\documentclass[journal]{IEEEtran}

\usepackage[utf8]{inputenc}
\usepackage[english]{babel}
\usepackage{booktabs}
\usepackage{multirow}
\usepackage{amssymb}
\usepackage[table,xcdraw]{xcolor}
\usepackage{cite}
\usepackage{float}
\usepackage{graphicx}
\DeclareGraphicsExtensions{.pdf,.png}

\usepackage{caption}
\usepackage{subcaption}

\usepackage{balance}
\usepackage{algorithm}
\usepackage{algorithmic}
\usepackage[cmex10]{amsmath}
\usepackage{mathtools}
\usepackage{cuted}

\newcommand{\lora}{\textsc{LoRa}}
\newcommand{\lorawan}{\textsc{LoRaWAN}}
\newcommand{\sigfox}{\textsc{SigFox}}
\newcommand{\nbiot}{\textsc{NB-IoT}}

\usepackage{soul}
\usepackage{soulutf8}
\usepackage{color}

\soulregister\lora7
\soulregister\lorawan7
\soulregister\sigfox7
\soulregister\nbiot7
\soulregister\zbn7
\soulregister\wsun7
\soulregister\cite7
\soulregister\ref7

\newcommand{\textchange}[1]{#1}

\begin{document}
\bstctlcite{IEEEexample:BSTcontrol}

\title{\lora{} Performance Analysis with\\ Superposed Signal Decoding}

\author{Jean~Michel~de~Souza~Sant'Ana,~\IEEEmembership{Student Member,~IEEE,}
        Arliones~Hoeller,~\IEEEmembership{Member,~IEEE,}
        Richard~Demo~Souza,~\IEEEmembership{Senior Member,~IEEE,}
        Hirley~Alves,~\IEEEmembership{Member,~IEEE,}
        and Samuel~Montejo-Sánchez,~\IEEEmembership{Member,~IEEE}
\thanks{J. M. S. Sant'Ana, A. Hoeller and H. Alves are with the Centre for Wireless Communications, University of Oulu, Finland.}
\thanks{A. Hoeller and R. D. Souza are with the Dept. of Electrical and Electronics Engineering, Federal University of Santa Catarina, Florianóplis, Brazil.}
\thanks{A. Hoeller is also with the Dept. of Telecommunications, Federal Institute for Education, Science, and Technology of Santa Catarina, São José, Brazil.}
\thanks{S. Montejo-Sánchez  is with Programa Institucional de Fomento a la I+D+i (PIDi), Universidad Tecnológica Metropolitana, Santiago, Chile.}
\thanks{Corresponding author: Jean.DeSouzaSantana@oulu.fi}
\thanks{This work was partially supported in Brazil by CNPq, PrInt CAPES-UFSC ``Automation 4.0'', INESC P\&D Brasil (F-LOCO,  Energisa, ANEEL PD-00405-1804/2018); in Finland by Academy of Finland, 6Genesis Flagship (Grant 318927), EE-IoT (Grant 319008), and FIREMAN (Grant 326301); in Chile by FONDECYT Postdoctoral (Grant 3170021) and FONDECYT Regular No. 1201893.}
\thanks{\copyright2020 IEEE. Personal use of this material is permitted. Permission from IEEE must be obtained for all other uses, in any current or future media, including reprinting/republishing this material for advertising or promotional purposes, creating new collective works, for resale or redistribution to servers or lists, or reuse of any copyrighted component of this work in other works.}}

\maketitle

\begin{abstract}
This paper considers the use of successive interference cancellation (SIC) to decode superposed signals in Long Range (\lora{}) networks.
We build over a known stochastic geometry model for \lora{} networks and include the effect of recovering colliding packets through SIC.
We derive closed-form expressions for the successful decoding of packets using SIC taking path loss, fading, noise, and interference into account, while we validate the model by means of Monte Carlo simulations.
Results show that SIC-enabled \lora{} networks improve worst-case reliability by up to $34\%$.
We show that, for at least one test scenario, SIC increases by $159\%$ the number of served users with the same worst-case reliability level.
\end{abstract}

\section{Introduction}

The Internet-of-Things demands machine-type communications (MTC) to serve massive numbers of devices with low energy consumption and reasonable reliability. The first two requirements are addressed by Low-Power Wide-Area (LPWA) technologies like Long Range Wide Area Network (\lorawan{}), \sigfox{}, and \nbiot{}. However, to achieve large-scale connectivity and low energy consumption, LPWA networks tend to replace complicated channel control by simpler medium access protocols, at the cost of reliability~\cite{Ayoub:CST:2019}.

The \lorawan{} protocol stack, which uses Long Range (\lora{}) as physical layer technology~\cite{LoRaAlliance:2018}, emerges as a promising LPWA network solution. Its openness facilitates and encourages adoption by both researchers and practitioners, however it may experience decreased reliability with increased numbers of users~\cite{Georgiou:WCL:2017,Mahmood:TII:2019}. As a solution to improve performance, several works exploit diversity relying on independent realizations of the wireless channel. For instance, time diversity is approached in the form of independent~\cite{Hoeller:ACS:2018} or coded~\cite{Marcelis:IoTDI:2017,Sanchez:WCL:2019} message replications for both \lorawan{} or general LPWA networks. These approaches have drawbacks. Message replication improves performance at the cost of increased energy consumption and network usage, flooding the network and reducing the number of supported users.



In a different direction, the combination of power-domain non-orthogonal multiple access (NOMA) and successive interference cancellation (SIC) is a key enabler of next generation spectrum-efficient communications~\cite{islam:CST:2017}. Moreover,  \lora{} features the capture effect, where one out of several simultaneously received signals can be decoded provided that its signal-to-interference ratio (SIR) is sufficiently high, above a given threshold~\cite{Semtech:2015}.
In this line, Noreen \textit{et al.}~\cite{Noreen:EWC:2018} exploit this fact and simulate the performance of \lora{} with SIC to recover the interfering signals, not only the one recovered by the capture effect. Their results show that SIC can increase network performance.
Rachkidy \textit{et al.}~\cite{Rachkidy:LCN:2018} investigate interference cancellation at the symbol level and propose two algorithms for implementing the technique.
Laport-Fauret \textit{et al.}~\cite{Laporte-Fauret:PIMRC:2019} and Xia \textit{et al.} \cite{Xianjin:SenSys:2019} present  techniques to decode superposed \lora{} signals taking advantage from the preamble and chirp characteristics of \lora{} technology to tune the signals and remove them separately.
However, common to \cite{Noreen:EWC:2018,Rachkidy:LCN:2018, Laporte-Fauret:PIMRC:2019,Xianjin:SenSys:2019} is the lack of closed-form network performance analysis. 


In this paper, we consider the effect of a SIC-enabled gateway, able to recover superposed packets when we have one interfering signal, on the performance of a \lora{} network. An important advantage of this technique is that network servers and end-devices may operate agnostic of SIC at the gateway, while the method does not require any information on network traffic or a feedback link. Also, our work intends to motivate the implementation of practical techniques previously stated that can decode multiple superposed \lora{} signals.
This paper builds over the \lora{} network stochastic geometry model originally proposed in \cite{Georgiou:WCL:2017}, extending it to account for SIC-based decoding. Then, we provide a new closed-form performance analysis model, which shows that both the number of nodes in coverage and the average reliability can be substantially increased compared to a regular \lora{} network.



\section{\lora{}}
\label{sec:lora}

\lora{} is a sub-GHz Chirp Spreading Spectrum (CSS) technology optimized for long-range and low-power transmissions~\cite{LoRaAlliance:2018}.
It uses quasi-orthogonal spreading factors (SF) that increase link budget and the number of virtual channels at the expense of prolonged Time on Air (ToA).
Table~\ref{tab:lora_sensi} presents the relation between SF, ToA, bit rate, and link budget.

\lorawan{} is the most used \lora{}-based protocol stack, with a star topology where end-devices reach one or more gateways through single-hop links.
Gateways have IP connections to network servers.
\lorawan{} controls channel access with ALOHA and exploits SFs to enable concurrent connections in quasi-orthogonal virtual channels.
\lorawan{} provides many configurations based on SF, which varies from 7 to 12, and bandwidth ($B$), usually 125 kHz or 250 kHz for uplink.



\begin{table}[t]
\centering
\caption{\lora{} uplink characteristics: payload of 9 bytes, $B = 125$~kHz, CRC and header mode enabled~\cite{Semtech:2015}.}
\label{tab:lora_sensi}
\begin{tabular}{@{}cccccc@{}}
\toprule
\textbf{SF} & \textbf{\begin{tabular}[c]{@{}c@{}}Time-on-Air\\ToA - ms \end{tabular}} & \textbf{\begin{tabular}[c]{@{}c@{}}Bit rate\\kbps\end{tabular}} & \textbf{\begin{tabular}[c]{@{}c@{}}Sensitivity\\dbm\end{tabular}} & \textbf{\begin{tabular}[c]{@{}c@{}}SNR threshold\\$q_i$ - dB\end{tabular}} & \textbf{\begin{tabular}[c]{@{}c@{}}Duty cycle*\\$p_i \times 10^{-6}$\end{tabular}}  \\ \midrule
7   & 41.22     & 5.47 & -123       & -6       & 45.8    \\ 
8   & 72.19     & 3.12 & -126       & -9       & 80.2    \\
9   & 144.38    & 1.76 & -129       & -12      & 160.4   \\
10  & 247.81    & 0.98 & -132       & -15      & 275.3   \\
11  & 495.62    & 0.54 & -134.5     & -17.5    & 550.7   \\
12  & 991.23    & 0.29 & -137       & -20      & 1101.4  \\ \bottomrule
\multicolumn{6}{c}{*duty cycle considering ToA and a message generation period of 15 minutes.}
\end{tabular}
\end{table}

\section{Baseline System Model}
\label{sec:model}

Following~\cite{Mahmood:TII:2019}, we consider a circular coverage region $\mathcal{V} \subseteq \mathbb{R}^2$ with radius $R$ meters and area $V = \pi R^2$, with $\bar{N}$ end-devices on average deployed uniformly.
We split the coverage region into six rings, each using a different SF, assumed to be orthogonal among them, being $\text{SF}_7$ allocated to nodes within the innermost ring, and $\text{SF}_{12}$ to nodes in the outermost ring. 
Each ring $i$ is bounded by inner ($l_{i-1}$) and outer ($l_i$) radii.
We assume that the reference\footnote{We use subscript ``$1$'' to denote the reference node under analysis.} node is $d_1$ meters from the gateway.
We model activity in each ring by a homogeneous Poisson Point Process (PPP) $\Phi_i$ with intensity $\alpha_i = 2p_i\rho V_i$, $\alpha_i > 0$, where $p_i$ is the duty-cycle of nodes in ring $i$, $\rho = \frac{\bar{N}}{V}$ is the spatial density, and $V_i = \pi(l_i^2 - l_{i-1}^2)$ is the area of ring $i$.
The average number of devices in ring $i$ is $\bar{N}_i = \rho V_i$.
All devices transmit in the uplink at random using ALOHA, with the same bandwidth $B$ and the same fixed transmit power $\mathcal{P}_t$.

The distance from node $k$ to the gateway at the origin is $d_k$ meters.
We model path loss as $g_k = \left (\frac{\lambda}{4 \pi d_k}  \right )^\eta$, where $\lambda=\frac{c}{f_c}$ is the wavelength, $c$ is the speed of light, $f_c$ is the carrier frequency, and $\eta$ is the path loss exponent.
The model also assumes Rayleigh fading $h_k$, thus, fading power is exponentially distributed, \textit{i.e.}, $|h_k|^2 \sim \textup{exp}(1)$.

If the reference \lora{} node transmits the signal $s_1$, the received signal at the gateway, $r_1$, is the sum of the attenuated transmitted signal, interference and noise,
\begin{equation}
    r_1 \!=\! \sqrt{\mathcal{P}_t g_1} h_1  s_1 \!+\! \sum\nolimits_{k\in\Phi_i} {\sqrt{\mathcal{P}_t g_k} h_k s_k} \!+\! w \label{eqn:r1},
\end{equation}
where $k$ includes the active nodes in the PPP but the reference node, $w$ is additive white Gaussian noise with zero mean and variance $\sigma_w^2$,  and $i$ is the ring the node at $d_1$ belongs to.

A node is in coverage if it is connected to the gateway (probability $H_1$) and there is no collision (probability $Q_1$), which considers the SINR according to \cite{Haenggi:Book:2012}. A collision takes place when simultaneous transmissions use the same SF, and the signal-to-interference ratio (SIR) is below the threshold $\gamma$. Thus, the coverage probability\footnote{Note that this is a lower bound, since $H_1$ and $Q_1$ are dependent due to fading \cite{Lim:CL:2018}. Moreover, as \lora{} specifies two different thresholds, for radio sensitivity and  the capture effect~\cite{Semtech:SX1272:2015}, the independent approach is convenient. Fortunately, this simplified approach allows us to considerably reduce the complexity of the analytical derivations without relevant changes to the results.} is \cite{Georgiou:WCL:2017, Mahmood:TII:2019}
\begin{align}
    C_1 \textchange{\approx} H_1 Q_1. \label{eqn:c1old}
\end{align}

\subsection{Connection Probability}

The connection probability $H_1$ depends on the distance between a node and the gateway.
A node is connected if the signal to noise ratio (SNR) at the gateway is above a threshold.
The connection probability is
$H_1 = \mathbb{P}[\textup{SNR} \geq q_i ~|~ d_1, i]$, 
where $q_i$ is the SNR reception threshold, from Table~\ref{tab:lora_sensi}, dependent on $\text{SF}_i$,
Therefore, as the instantaneous $\textup{SNR} = \frac{\mathcal{P}_t |h_1|^2 g_1}{\sigma_w^2}$, $H_1$ is
\begin{equation}
    H_1 = \mathbb{P} \left [ |h_1|^2 \geq \frac{\sigma_w^2 q_i}{\mathcal{P}_t g_1} ~\middle|~d_1 \right ] = \textup{exp} \left ( - \frac{\sigma_w^2 q_i}{\mathcal{P}_t g_1} \right ) \label{eqn:h1_2}.
\end{equation}

\subsection{Capture Probability}

As in~\cite{Mahmood:TII:2019}, we model the SIR as
\begin{align}
    \textup{SIR} &= \frac{\mathcal{P}_t |h_1|^2 g_1}{\sum_{k\in\Phi_i} \mathcal{P}_t |h_k|^2 g_k} = \frac{|h_1|^2 d_1^{-\eta}}{\sum_{k\in\Phi_i} |h_k|^2 d_{k}^{-\eta}}.\label{eqn:sir}
\end{align}
The probability of a successful reception in the presence of interference, considering the capture effect, is
\begin{align}
    Q_1 &= \mathbb{P}\left[\textup{SIR} \geq \gamma ~|~ d_1 \right],
\end{align}where $\gamma$ is the capture threshold.
As detailed in~\cite{Hoeller:ACS:2018,Mahmood:TII:2019}, $Q_1$ is
\begin{align}
    Q_1 = \textup{exp} \Biggr\{ -\pi & \frac{\alpha_i}{V_i} \left[ l_i^2 ~ _2F_1 \left(1,\frac{2}{\eta};1+\frac{2}{\eta}; -\frac{l_i^{-\eta}}{\gamma d_1^\eta} \right) \right. \ldots \nonumber \\
    &- \left. l_{i-1}^2 ~ _2F_1 \left(1,\frac{2}{\eta};1+\frac{2}{\eta}; -\frac{l_{i-1}^{-\eta}}{\gamma d_1^\eta} \right) \right] \Biggr\}, \label{eqn:q1}
\end{align}
where $_2F_1(\cdot)$ is the Gauss Hypergeometric function and $l_{i-1}$ and $l_i$ are the boundaries of the $\text{SF}_i$ ring the node at $d_1$ is.

\section{Coverage Probability with SIC}
\label{sec:sic}

According to~\cite{Georgiou:WCL:2017}, \lora{} capture probability presents the near-far problem, where transmissions from nodes further away from the receiver are suppressed by transmissions of nodes closer to the gateway due to path loss.
Therefore, the further away the node is from the gateway, the less likely it is that it benefits from the capture effect.
However, if the receiver can decode the stronger received message from the colliding messages, by performing SIC it may be able to decode a weaker message too.
Although, in theory, SIC could be used iteratively to decode several colliding messages, it is unlikely to happen unless transmissions are carefully planned to that end, like in power-domain NOMA systems~\cite{lopez:TWC:2018}.
Fortunately, the case where only two simultaneous signals arrive at the gateway is very common in \lora{} networks, as shown later in Section~\ref{sec:results}.
Moreover, considering only a single SIC iteration reduces the receiver complexity and decoding latency.
Therefore, we consider to use SIC in cases where there are two colliding transmissions,
trying to recover both messages. 

\begin{figure}[tb]
    \centering
    \includegraphics[width=\columnwidth]{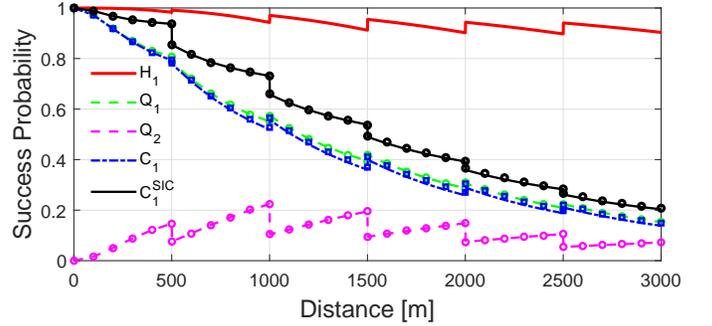}
    \caption{Theoretical (lines) and simulated (markers) probabilities as a function of the distance from the gateway ($d_1$).}
    \label{fig:simul}
    \vspace{-8mm}
\end{figure}

The coverage probability of the signal of interest accounting for SIC becomes\footnote{\textchange{Since we are mixing a lower bound (considering $H_1$, $Q_1$ and $Q_2$ to be independent) with the upper bound $H_2 \leq H_1$, we can only say that this is an approximation, which is better seen in Figure~\ref{fig:simul} and discussed in Section \ref{sec:results}.}}
\begin{align}
    C_1^{\textup{SIC}} \approx H_1Q_1 + H_1Q_2, \label{eqn:c1new}
\end{align}
where the first term ($H_1Q_1$) is $C_1$ as in~\eqref{eqn:c1old}, \textit{i.e.}, the probability that the signal of interest is decoded because it is sufficiently stronger than noise and interference.
The second term ($H_1Q_2$) considers decoding the interfering signal first. Then, we apply SIC and decode the signal of interest without interference once it is sufficiently stronger than noise.

In order to derive the second part of \eqref{eqn:c1new}, first consider $G_j = |h_j|^2g(d_j)P_t$ as the channel gain of signal $j$, $j\in\{1,2\}$\footnote{We use subscript ``$2$'' to denote the interfering node, when we apply SIC.}. Then, we can write the probability $H_2$ that the signal of interest $s_1$ and the interfering signal $s_2$ are in connection (thus sufficiently stronger than noise), given that $s_2$ can be captured, as
\begin{align}
    H_2 &= \mathbb{P}\left[G_2 \geq q_i\sigma_w^2~\cap~G_1 \geq q_i\sigma_w^2 ~|~ G_2 \geq \gamma G_1, ~d_1,~i\right] \nonumber \\
    &= \mathbb{P}\left[G_1 \geq q_i\sigma_w^2 ~|~ G_2 \geq \gamma G_1,~d_1\right] \textchange{\approx} ~H_1. \label{eqn:approx_h2}
\end{align}\indent Moreover, the probability that the interfering signal is sufficiently stronger than the reference signal, so that it can be decoded by the capture effect, is 
\begin{align}
    Q_2 &\stackrel{(a)}{=}\mathbb{P}\left[\frac{G_2}{G_1}\geq \gamma~\middle|~|\Phi_i|=1,~d_1\right] \nonumber \\
    &\stackrel{(b)}{=}\mathbb{P}\left[ |h_2|^2\geq \frac{\gamma|h_1|^2d_1^{-\eta}}{D_2^{-\eta}} ~\middle|~d_1 \right] \mathbb{P}\left[ |\Phi_i| = 1 \right] \nonumber \\
    &\stackrel{(c)}{=}\mathbb{E}_{D_2}\left[ \frac{d_1^{\eta}}{d_1^{\eta} + \gamma D_2^{\eta}} \right] \alpha_i e^{-\alpha_i} , 
\end{align}
where in $(a)$ we consider only one interfering node, and in $(b)$ we decondition from the number of interfering nodes. $D_2$ is a random variable (RV) describing $d_2$. In $(c)$ we average over the fadings $|h_j|^2\!\sim\!\text{exp}(1)$ and the cardinality $|\Phi|$, which is a Poisson RV of mean $\alpha_i$ and probability density function (PDF) $f_{|\Phi_i|}(k) = \frac{\alpha_i^k e^{-\alpha_i}}{k!}$, where $k\!=\!1$ is the number of interferers.

Then, since nodes are distributed in $\mathcal{V} \subseteq \mathbb{R}^2$, $f_{D}(d)=\frac{2d}{R^2}, 0\leq d\leq R$~\cite[eq. (3)]{Bharucha:RLC:2008}.
We generalize the ring-based geometry PDF as $f_{D_2}(d_2) = \frac{2d_2}{l_i^2 - l_{i-1}^2}, l_{i-1} \leq d_2 \leq l_i$, which allows us to take the expectation over $D_2$ and obtain 
\begin{align}
    Q_2 &= \frac{\alpha_i e^{-\alpha_i}2d_1^\eta}{l_i^2 - l_{i-1}^2}\int_{l_{i-1}}^{l_i} \frac{x}{d_1^\eta + \gamma x^\eta} \textup{d}x.\label{eqn:q2expect}
\end{align}Rearranging~\eqref{eqn:q2expect}, using the Pochhammer function and the Gaussian Hypergeometric function~\cite{Daalhuis:Chapter:2010}, we have that
\begin{align}
    Q_2 = \frac{\alpha_i e^{-\alpha_i}}{l_i^2-l_{i-1}^2} \left[ 
        l_i^2 \,_2F_1\left( 1, \frac{2}{\eta}; \frac{2}{\eta} + 1; -\frac{\gamma l_i^\eta}{d_1^\eta} \right) \ldots \right. \nonumber \\
        \left. - l_{i-1}^2 \,_2F_1\left( 1, \frac{2}{\eta}; \frac{2}{\eta} + 1; -\frac{\gamma l_{i-1}^\eta}{d_1^\eta} \right)
    \right].\label{eqn:q2}
\end{align}

\begin{figure}[tb]
    \centering
    \includegraphics[width=1\columnwidth]{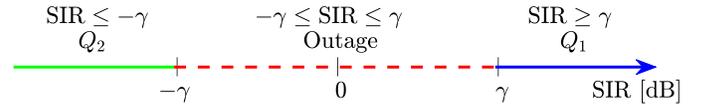}
    \caption{Graphical representation of capture probabilities $Q_1$ and $Q_2$ as a function of the SIR of the signal of interest.}
    \label{fig:sir_range}
    \vspace{-5mm}
\end{figure}

\textchange{Finally, $Q_1$ and $Q_2$ represent probabilities of disjoint events, as illustrated in Figure \ref{fig:sir_range}, thus they can not happen simultaneously.}  When the SIR is above $\gamma$, the signal of interest can be decoded due to the capture effect. However, if the SIR is below $-\gamma$, and there is only one interfering signal, it means that this interfering signal SIR is above $\gamma$, and thus it can be decoded first, due to the capture effect.

\section{Numeric Results}
\label{sec:results}

We evaluate the impact of SIC in \lora{} networks in terms of reliability. Unless stated otherwise, we consider $\gamma=1$dB, $\eta=2.8$, $f_c=868$MHz, $\mathcal{P}_t=14$dBm, $B=125$kHz, and $\sigma_w^2 = -174+F+10\log_{10}B = -117$dBm with receiver noise figure $F=6$dB. Inspired by~\cite{Georgiou:WCL:2017,Hoeller:ACS:2018,Mahmood:TII:2019}, we define the deployment area with a fixed radius of $R=3000$m, and divide the area in six rings with same width $l_i - l_{i-1}=500$m.
Nodes in the innermost ring use $\text{SF}_7$, while SF increases with $i$. Moreover, the duty cycle is $p_i=1\%, \forall i$. This setup configures a typical suburban scenario following European regulations.

In Figure~\ref{fig:simul} we seek to validate the proposed models by comparing the theoretical expressions to simulations.
In the figure, lines represent the expressions in~\eqref{eqn:c1old}, \eqref{eqn:h1_2}, \eqref{eqn:q1}, \eqref{eqn:c1new}, and \eqref{eqn:q2}, while marks show the average of $10^5$ Monte Carlo simulations \textchange{considering all dependencies between the events.}
Note that the approximation in \eqref{eqn:approx_h2} is good as the Monte Carlo simulations of $C_1^{\textup{SIC}}$ match very well with our proposed framework.
For the worst-case scenario scenario in the figure, \textit{i.e.}, for the node at $d_1=R=3000$m, $H_1$ shows that around $10\%$ of messages are lost due to disconnection, and around $80\%$ are lost due to collisions ($Q_1$), resulting in coverage probability ($C_1$) around $13\%$.
However, if SIC for one interfering message is considered, as in this paper, the collision probability drops to around $22\%$ ($Q_1 + Q_2$), resulting in a coverage probability with SIC ($C_1^{\textup{SIC}}$) around $20\%$.
Looking to the $Q_2$ curve, we notice that SIC decoding probability grows with the distance within the SF ring. That is because if the reference signal is closer to the ring outer border, it is likely to be weaker than the interfering signal, what increases the chance of the interfering signal being successfully decoded, thus allowing the reference  signal to be decoded too.

\begin{figure}
\centering
\begin{subfigure}{.5\columnwidth}
  \centering
  \includegraphics[width=\linewidth]{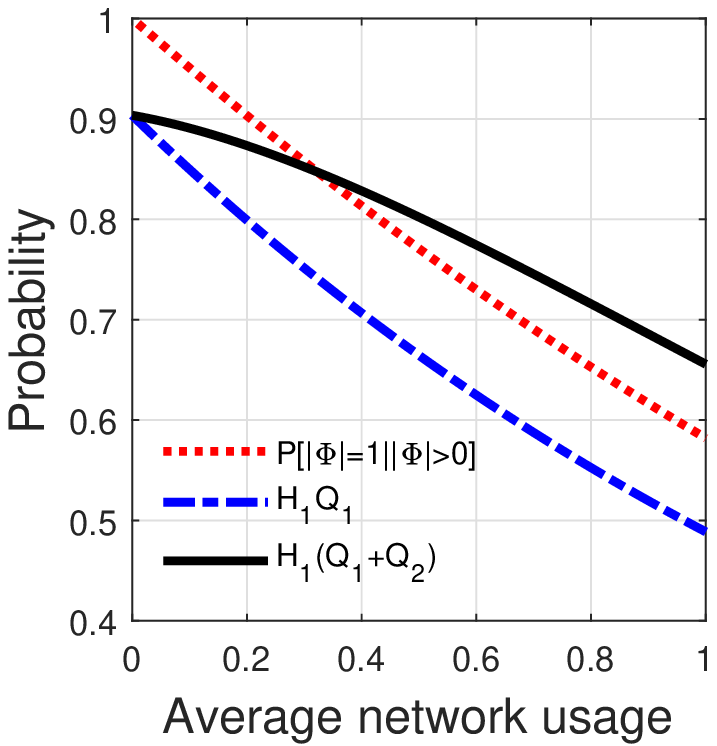}
  \caption{Varying $\alpha_i$, $\gamma=1$dB.}
  \label{fig:alpha_range}
\end{subfigure}%
\begin{subfigure}{.5\columnwidth}
  \centering
  \includegraphics[width=\linewidth]{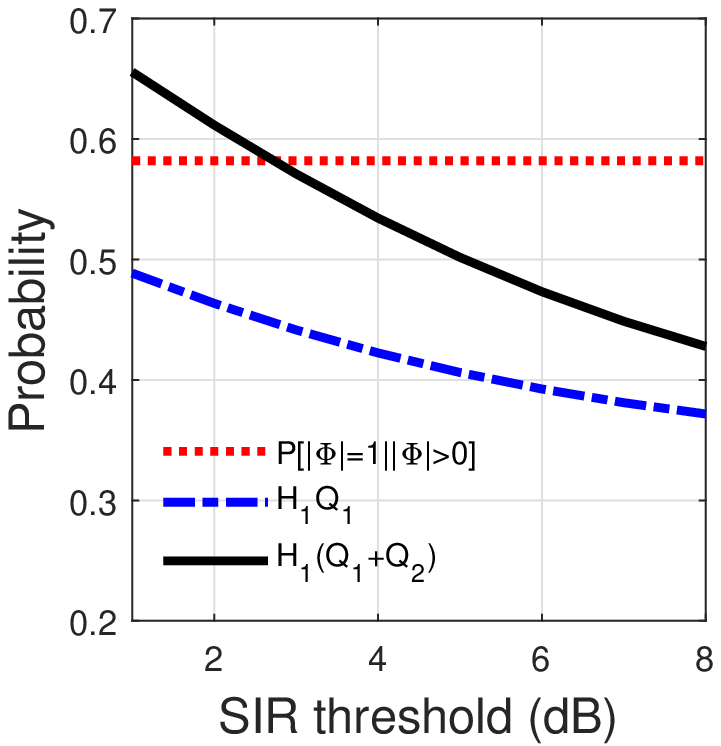}
  \caption{$\alpha_i=1$, varying $\gamma$.}
  \label{fig:gamma_range}
\end{subfigure}
\vspace{-5mm}
\caption{Probabilities  at the cell border ($d_1=3000$~m).}
\label{fig:ranges}
\vspace{-8mm}
\end{figure}

Figure~\ref{fig:ranges} considers the performance of the worst-case node in Figure~\ref{fig:simul}, that at the cell border ($d_1=3000$m).
Plot~\ref{fig:alpha_range} shows capture probabilities with and without SIC, \textit{i.e.}, $H_1(Q_1+Q_2)$ and $H_1Q_1$, respectively, as a function of network usage rate $\alpha_i$.
Plot~\ref{fig:gamma_range} shows the same capture probabilities as a function of the SIR threshold $\gamma$ with fixed $\alpha_i=1$.
In both plots, the dotted/red curve shows $\mathbb{P}[|\Phi_i|=1~|~|\Phi_i|>0]$ -- the probability that there is only one interfering signal, provided there is a collision. For instance, when the usage rate is at $100\%$ ($\alpha_i=1$), $58.2\%$ of the collisions have only one interferer.
Plot~\ref{fig:alpha_range} shows that the gain is significant and increases with network usage, \textit{i.e.}, denser networks benefit more from SIC.
Note that in Figure~\ref{fig:alpha_range} the x-axis is equivalent to varying $N_i$ from $0$ to $100$ nodes because $N_i=\frac{\alpha_i}{p_i}$ and $p_i=1\%,\forall i$.
Plot~\ref{fig:gamma_range} shows that higher SIR thresholds decrease the SIC benefit, although $H_1Q_2=5.6\%$ for $\gamma=6$dB.
\lora{} transceiver manufacturer reports $\gamma=6$dB, although recent results show that it can be as low as $\gamma=1$dB~\cite{Mahmood:TII:2019, Croce_CL_2018}, where $H_1Q_2=16.7\%$.

\begin{table}[t]
    \centering
    \caption{Number of nodes using the duty-cycles in Table~\ref{tab:lora_sensi}.}
    \begin{tabular}{cccccccc} \toprule
        $\alpha_i$ & $\text{SF}_{7}$ & $\text{SF}_{8}$ & $\text{SF}_{9}$ & $\text{SF}_{10}$ & $\text{SF}_{11}$ & $\text{SF}_{12}$ & Total \\ \midrule
        0.20 & 2,183 & 1,247 & 623 & 363 & 182 & 91 & 4,689 \\
        0.52 & 5,677 & 3,241 & 1,621 & 944 & 472 & 236 & 12,191  \\
        1 & 10,917 & 6,233  & 3,116 & 1,815 & 907 & 454 & 23,442 \\ \bottomrule
    \end{tabular}
    \label{tab:Nxp_tradeoffs}
    \vspace{-5mm}
\end{table}

Since $\alpha_i = 2p_i\rho V_i=2p_iN_i$ and we fix $p_i$ and $V_i$, the results in Figure~\ref{fig:alpha_range} can be extrapolated  trading-off duty cycle ($p_i$) for number of nodes ($N_i$).
This is shown in Table~\ref{tab:Nxp_tradeoffs} considering the ToA and duty cycles from Table~\ref{tab:lora_sensi}.
Results in Table~\ref{tab:Nxp_tradeoffs} consider all nodes running the same application.
If $\alpha_i=1$ and $\gamma=1$, the network would support $23,442$ nodes with worst-case packet loss due to interference of $51.1\%$ without SIC and $34.4\%$ with SIC.
Moreover, suppose one wants to plan the network for a maximum loss of $20\%$.
From Figure~\ref{fig:alpha_range}, $H_1Q_1 > 0.8$ for $\alpha_i<0.20$ and $H_1(Q_1+Q_2) > 0.8$ for $\alpha_i<0.52$.
Then, Table~\ref{tab:Nxp_tradeoffs} shows that the use of SIC increases the number of nodes from 4,689 to 12,191 ($2.59\times$). 

\section{Conclusions}
\label{sec:conclusion}

We proposed a stochastic model extension for \lora{} networks considering SIC at the gateway, with at most two colliding packets. Such assumption was shown to be representative of most of collision events. The proposed method significantly increases the \lora{} network performance in terms of coverage probability, while such benefit comes at no cost at the nodes, as the SIC  processing is performed only at the gateway. In future work, we aim to maximize the success probability by transmit power allocation that maximizes $Q_1+Q_2$. 


\bibliographystyle{IEEEtran}
\bibliography{references}

\end{document}